# Speech Emotion Recognition Using MFCC Features and LSTM-Based Deep Learning Model

Adelekun Oluwademilade[1*], Ademola Adedamola[1,2], Abiola Abdulhakeem[1,3], Akinpelu Azeezat[1,4], Eraiyetan Israel[1,5], Omotosho Oluwadunsin[1,6], Ibenye Ikechukwu[1,7], Ayuba Muhammad[2], Olusanya Olamide O.[3], Kamorudeen Amuda[4]

[1*] Department of Electrical Electronics and Telecommunications Engineering, Bells University of Technology, P.M.B. 1015, Ota, Ogun State, Nigeria
oluwademiladeadelekun@gmail.com
[2] New Horizons Bells University of Technology, P.M.B. 1015, Ota, Ogun State, Nigeria
[3] Department of Computer Engineering, Bells University of Technology, P.M.B. 1015, Ota, Ogun State, Nigeria
[4] Department of Computer and Information Sciences, Towson University, Maryland, USA

***Abstract-*** Speech Emotion Recognition (SER) is the use of machines to detect the emotional state of humans based on the speech, which is gaining importance in natural human-computer interaction. Speech is a very valuable source of information, as emotions modify the patterns of speech; pitch, energy and even timing. Nonetheless, SER is not an easy task because speakers are not constant, and situations vary when recording and the sound similarity between specific feelings. In this work, the author introduces a speech emotion recognition system relying on the Mel-Frequency Cepstral Coefficient and Long Short-Term Memory (LSTM) neural network, as a feature extraction method. The Toronto Emotional Speech Set (TESS) speech signal was pre-processed, and transformed into MFCC features to understand the important aspects in terms of time. The resultant features were then introduced to LSTM model, which is able to learn long term features of sequential audio data. The trained model was measured over several emotion classes occurring in the dataset. As seen in the results of experiments, the proposed MFCC-LSTM approach succeeds in capturing the patterns of emotions in speech and provides highly realistic classifications in all the chosen emotion classifications. This study presents a speech emotion recognition system using Mel-Frequency Cepstral Coefficients (MFCCs) as features and a deep learning LSTM classifier. A Support Vector Machine (SVM) with an RBF kernel served as a classical baseline, achieving 98% accuracy, against which the proposed LSTM model, achieving 99% accuracy, was validated. Overall, it is possible to confirm that LSTM-based architectures can be used to address the task of speech emotion recognition. Actual applications of the proposed system may be virtual assistants and mental health surveillance.

## 1. INTRODUCTION

Human speech carries more than the words, it sends the feelings in a form of different tones, pitch, rhythm, and intensity. Such vocal gestures are crucial in the interpersonal communication, which define interpretation and the level of socialization. The majority of the computational systems, however, are mostly concerned with the recognition of the spoken material without paying any attention to the emotive meaning of the speech. The drawback lessens the efficiency of intelligent systems including virtual assistant, mental health monitoring, automated customer services, and education technologies, in which emotional awareness would progress responsiveness and personalization. Speech Emotion Recognition (SER) aims to fill this gap and allow machines to identify the emotional condition based on speech data only [3], [14]. Good SER systems are applied in such domains as human-computer interaction, mental health monitoring, adaptive learning and customer service analytics [14], [15]. These systems can adjust their behaviour to suit the requirements of users and make interactions more natural and high-quality by detecting emotion, e.g. happy/sad/angry/neutral, and helping them react appropriately. Historically, SER used handcrafted acoustic characteristics, such as Mel-Frequency Cepstral Coefficients (MFCCs), pitch, energy, formants and prosodic cues [17], [18]. They were typical inputs to classical machine learning classifiers (Support Vector Machines, or SVMs, and k-Nearest Neighbors, k-NN) [3], [5]. With the introduction of deep learning, and specifically the Long Short-Term Memory (LSTM) network, models finally were able to automatically learn temporal correlations in speech, which are important in emotion recognition [5], [15]. LSTMs have the ability to represent changes in emotional patterns that take place during an utterance over time which has a better accuracy in comparison to conventional classifiers [6], [8]. The network used in this work is a unidirectional LSTM network that was trained based on Toronto Emotional Speech Set database (TESS) to recognize seven categories of emotions. The strategy is based on the concept of identifying MFCC characteristics in every utterance and carefully predicting the emotional condition with the help of time modeling [8]. The rest of this paper will be structured in the following way. In Section 2, the TESS dataset, preprocessing and LSTM-based methodology have been identified. Section 3 contains the results of experiments and evaluation measures and Section 4 finalizes the paper, including the possible applications and the future directions of research.

## 2. METHODOLOGY

### 2.1 Dataset Description

We used the Toronto Emotional Speech Set (TESS) dataset which was employed to train and evaluate the proposed Speech Emotion Recognition system. TESS is a widely used benchmark dataset designed for controlled emotion recognition research. It consists of speech recordings produced by two professional female actors, each portraying a range of emotional states through predefined spoken utterances.

The Toronto Emotional Speech Set (TESS) dataset was used in this study. The original TESS dataset contains 2,800 audio recordings produced by two professional female actors, each uttering 200 target words under seven emotional conditions: angry, disgust, fear, happy, pleasant surprise, sad, and neutral.

Each emotion is equally represented, ensuring a balanced class distribution and reduced bias during model training and evaluation.

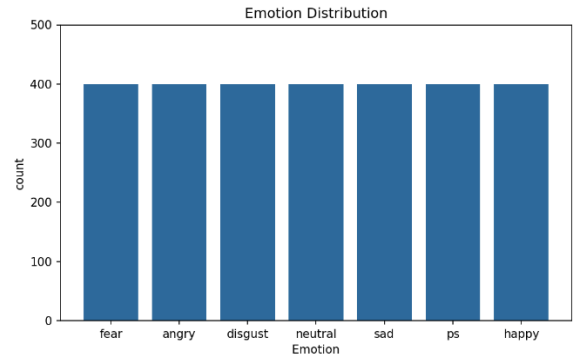

**Fig 1:** Emotion Count

The dataset was accessed via Kaggle for academic research purposes., and all recordings are anonymized. As a result, no additional ethical approval was required. The data were used strictly for academic research.

### 2.2 Data Preprocessing

Prior to model training, the speech data underwent several preprocessing steps to ensure consistency and effective feature representation. Each audio file in the dataset was first organized into a structured format by pairing the raw audio samples with their corresponding emotion labels. This resulted in a tabular representation where each entry consisted of an audio signal and its associated emotional class.

Exploratory data analysis (EDA) was then conducted to gain insights into the sound characteristics of the different emotional categories. Using the **Librosa library**, audio signals were loaded and visualized through waveform plots, illustrating amplitude variations over time. Spectrograms were also generated to examine frequency content across emotions, allowing qualitative observation of temporal and spectral differences between emotional states.

For Happy,

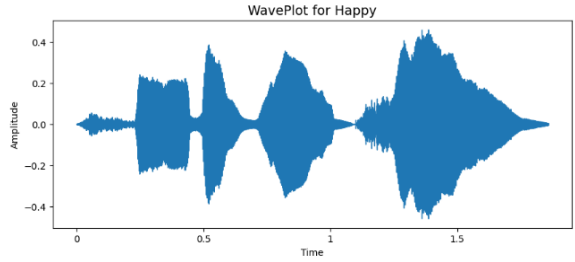

**Fig 2:** Waveplot for Happy Emotion

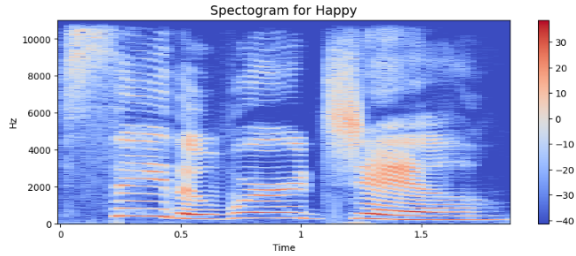

**Fig 3:** Spectogram for Happy Emotion

For Pleasant Surprise (ps),

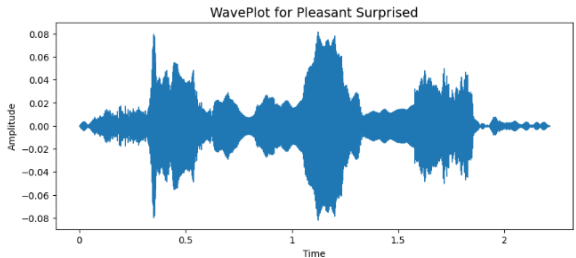

**Fig 4:** Waveplot for Pleasant Surprised Emotion

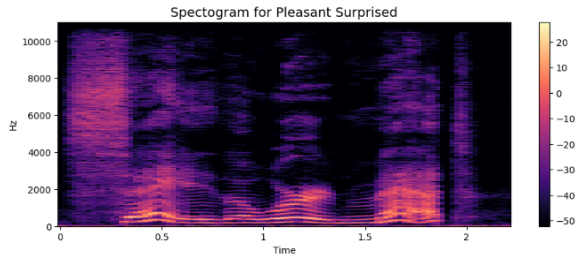

**Fig 5:** Spectogram for Pleasant Surprised Emotion

For Fear,

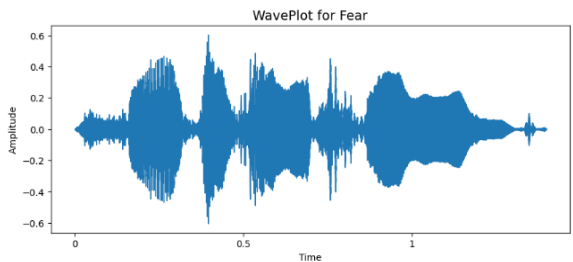

**Fig 6:** Waveplot for Fear Emotion

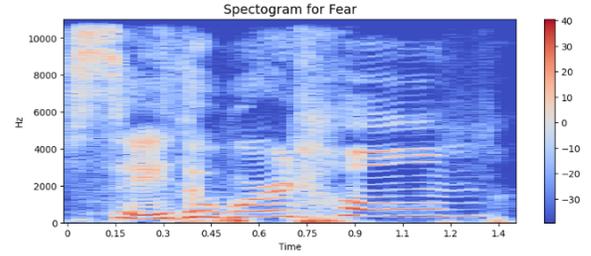

**Fig 7:** Spectogram for Fear Emotion

For Sad,

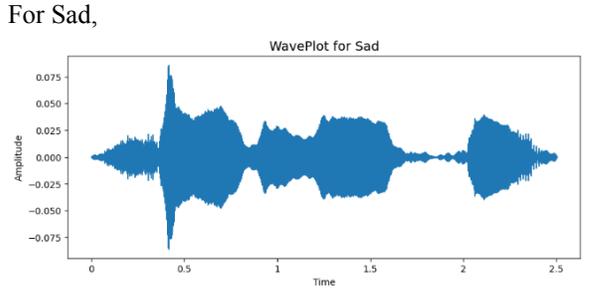

**Fig 8:** Waveplot for Sad Emotion

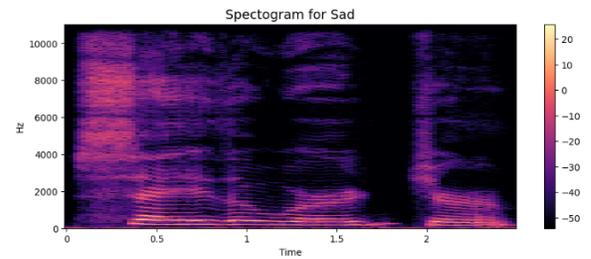

**Fig 9:** Spectogram for Sad Emotion

For Disgust,

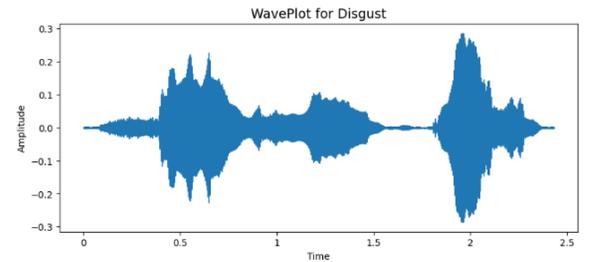

**Fig 10:** Waveplot for Disgust Emotion

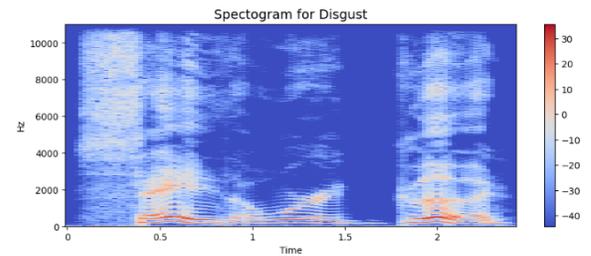

**Fig 11:** Spectogram for Disgust Emotion

For Angry,

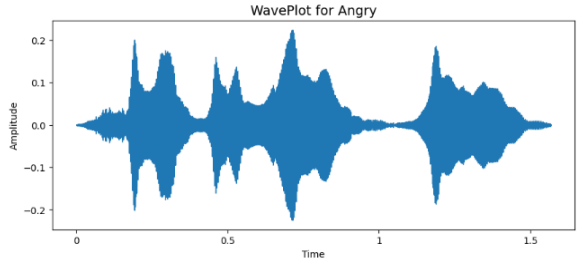

**Fig 12:** Waveplot for Angry Emotion

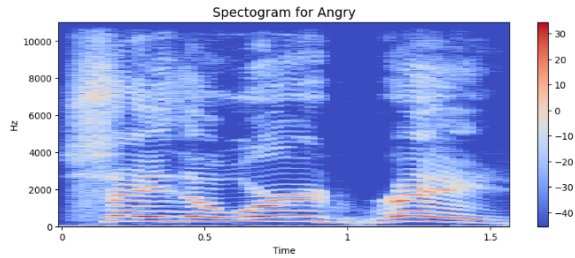

**Fig 13:** Spectogram for Angry Emotion

For Neutral,

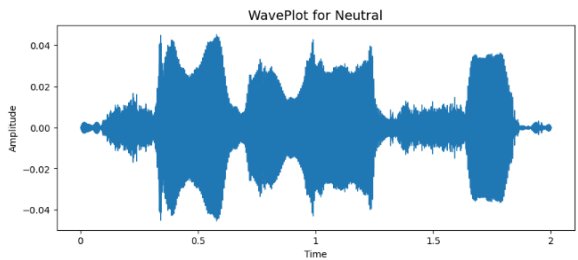

**Fig 14:** Waveplot for Neutral Emotion

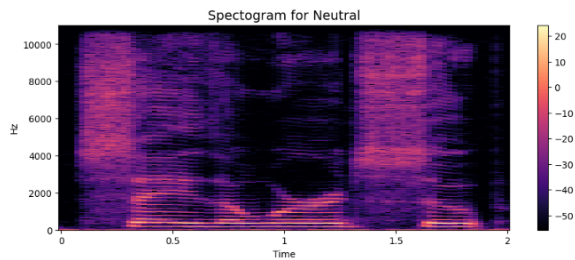

**Fig 15:** Spectogram for Neutral Emotion

Following exploration, feature extraction was done using Mel-Frequency Cepstral Coefficients (MFCCs), which are widely used in speech processing due to their ability to capture relevant spectral information [17], [18]. All audio files were trimmed to 3 seconds using librosa.load with duration=3 to ensure a fixed-length time dimension. MFCCs were extracted using a sampling rate of 22,050 Hz, n_fft=2048, hop_length=512, and 40 MFCC coefficients. Each utterance was represented as a sequence of 40-dimensional MFCC vectors across *t* time steps. The resulting input tensor had shape (N, t, 40), where N is the number of samples in order to fit the needs of the Long Short-Term Memory (LSTM) network [8]. One-hot encoding turned emotion labels into categorical representations, which helped the model learn how to classify emotions into more than one class.

This preprocessing pipeline created a consistent and useful representation of speech signals that could be used with LSTM architectures for sequential modeling.

### 2.3 Model Architecture

#### 2.3.1. Overview of the Model

In the given study, a Speech Emotion Recognition (SER) system, which is grounded on a Long Short-Term Memory (LSTM) neural network, is proposed. It is a model that is developed to capture the temporal patterns occurring in speech that are closely related to emotional expression. As there is time variation in the speech process via changing of pitch, intensity, and rhythm, the LSTM networks are best fitted to complete such a task because the networks handle sequential data [6], [8].

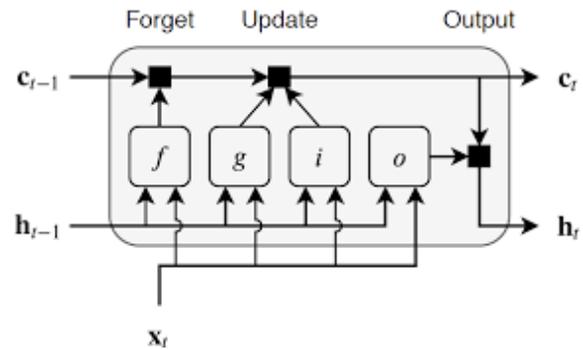

**Fig 16:** Block Diagram of LSTM Model

In contrast to more complicated hybrid architecture, the presented proposal assumes a single-stream LSTM architecture. This choice of design assists in minimizing the complexity of the model but high emotional representations are trained using speech in an effective manner [4], [9], [10]. The system works only with Mel-Frequency Cepstral Coefficient (MFCC) features as they are obtained out of audio utterances.

#### 2.3.2. LSTM-Based Feature Modeling

In the case of every speech utterance, MFCC features are calculated and arranged as a time series, with a single time component representing a short amount of time of the audio signal. In this paper, spectral features of the speech that are application in emotion recognition are represented by 40 MFCC features.

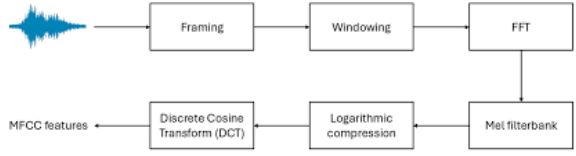

**Fig 17:** MFCC Feature Extraction Block Diagram

As an input in a unidirectional LSTM layer, the MFCC sequences are inserted. LSTM processes the sequence sequentially, by selectively retaining information that is emotional and ignoring the less significant variations.

Standard recurrent networks struggle with long sequences because gradients either vanish or blow up during backpropagation, making it difficult to retain information across many time steps. The LSTM addresses this through a dedicated memory cell, $C_t$, alongside three gating mechanisms that together control what the network keeps, discards, or passes forward at each step $t$ [19].

Given the input vector $x_t$ and the previous hidden state $h_{t-1}$, the network first decides what to throw away. The forget gate $f_t$ produces a value between 0 and 1 for each element of the cell state, where values near zero mean "discard" and values near one mean "keep":

$$f_t = \sigma(W_f[h_{t-1}, x_t] + b_f)$$

At the same time, the input gate $i_t$ and a candidate cell state $\tilde{C}_t$ work together to determine what new information gets written in:

$$i_t = \sigma(W_i[h_{t-1}, x_t] + b_i)\quad \tilde{C}_t = tanh(W_c[h_{t-1}, x_t] + b_c)$$

The cell state is then updated by blending what was retained from before with whatever new content the input gate admits:

$$C_t = f_t \odot C_{t-1} + i_t \odot \tilde{C}_t$$

Where $\odot$ is element-wise multiplication. This additive update is what makes LSTMs resistant to the vanishing gradient problem, as the cell state can carry information across many time steps with minimal modification.

Finally, the output gate $o_t$ regulates how much of the cell state feeds into the hidden state $h_t$, which is what gets passed to the next time step and ultimately used for classification:

$$o_t = \sigma(W_o[h_{t-1}, x_t] + b_o) \quad h_t = o_t \odot tanh(C_t)$$

In all equations, $W_f, W_i, W_c, W_o$ are learned weight matrices and $b_f, b_i, b_c, b_o, b_f, b_i, b_c, b_o$ are bias terms.

The sigmoid function σ constrains gate outputs to (0,1), while $tanh$ maps cell state values to (-1,1) before they influence the hidden state.

For this study, this sequential processing made LSTM a natural fit for the task. MFCC features extracted from the TESS dataset are inherently time-series data, and emotional cues in speech such as shifts in pitch, energy, and rhythm unfold across frames rather than appearing at a single point. The forget and input gates in particular allow the model to weight emotionally salient frames more heavily while discarding background noise or neutral phonetic content.

This allows the model to represent long term dependencies and time dynamics projected in emotional speech. The LSTM layer output is a tiny representation of the emotional content of the whole utterance.

### 2.3.3. Output Layer and Emotion Classification

The resulting representation produced by the LSTM layer is then sent to a fully connected (dense) layer that does emotion classification. This layer projects the learnt temporal features to an established category of emotions. The output layer has a softmax activation function to generate a probability distribution amongst the classes of emotion. The system that has the most probable emotion chosen is sent as the system output.

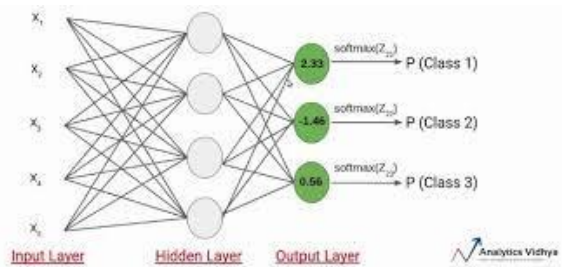

**Fig 18:** Softmax Activation Function

One-hot encoding is applied to the labels in emotion space and categorical cross-entropy is taken as the loss in training to facilitate the process of learning based on supervision. In this arrangement, the model is able to learn to discriminate features of emotions.

### 2.3.4. Model Design Considerations

The given LSTM-only model is focused on simplicity, efficiency, and brevity. With MFCC features and temporal modeling based only on LSTM, the system cannot incur the extra computational cost incurred by the convolutional/or multi-branch models. Such streamlined design renders the model applicable to the practical use like the real-time emotion-aware system, voice-based interface, and the affective computing tools. Although it is simple, the architecture can learn relevant emotion-related patterns of speech signals.

### 2.4 Training Procedure

The model of the LSTM-based speech emotion recognition was trained in a supervised learning environment, using speech samples from the TESS dataset. The dataset was partitioned using a stratified 80/20 train-test split, ensuring proportional representation of all seven emotion classes in both subsets. When processed with feature extraction and pre-processing, individual audio utterances were transformed into a three-dimensional input tensor, with dimensions matching the count of MFCC features and time-points, which can be sequentially modeled. This network was realized based on a sequential architecture with an input layer, a single layer of LSTM, and multiple fully connected layers. The LSTM layer had 128 hidden units and was set to produce one output representation sequence of an utterance. This representation was then given to dense layers having rectified linear unit (ReLU) activation functions in order to add non-linearity to learning. The dense layer was directly followed by dropout layers whose dropout rate is 0.2 to avoid overfitting. The last output layer made use of the softmax activation function to categorize every audio sample into one out of seven categories of emotions. Categorical cross-entropy loss amount was the model training and appropriate when compared to classification problems involving more than a single classification. Adam, the optimization method was adopted on the factor that it has an adaptive learning rate and has been proven to be effective with regard to training deep neural networks. Classification accuracy was used as a measure to determine model performance. The training was done on a 100 epochs and a batch size of 512 where the training data were shuffled at each epoch to avoid learning bias with ordered data. The accuracy of the validation and the loss were tracked during the training process so that the convergence behavior and the possibility of overfitting could be observed.

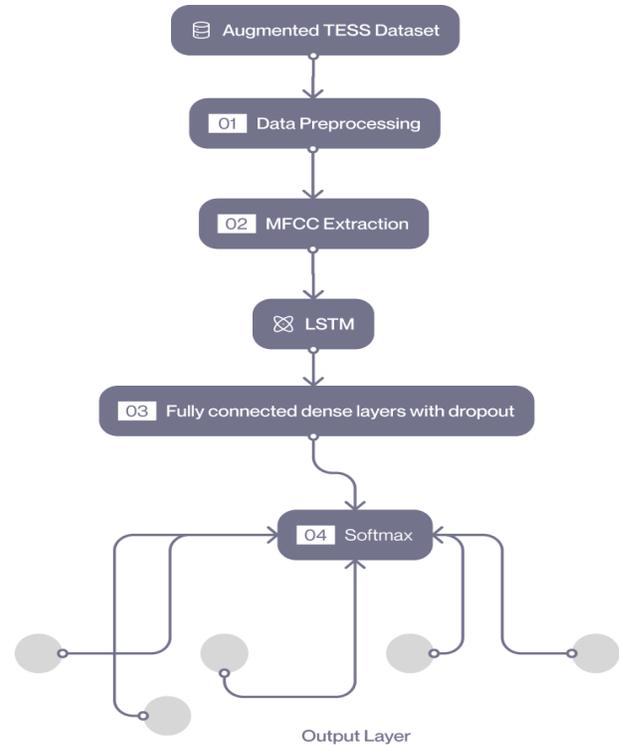

**Fig 19:** LSTM Architecture of the Project

### 2.4.1 SVM Baseline

To validate the performance of the proposed LSTM model, a Support Vector Machine (SVM) classifier with a Radial Basis Function (RBF) kernel was implemented as a classical baseline [3], [5]. Rather than operating on temporal MFCC sequences, the SVM receives a fixed-length feature vector computed by averaging the MFCC values across the time dimension for each utterance, producing a static 40-dimensional representation per sample. Prior to classification, these features were normalized using StandardScaler to zero mean and unit variance. The SVM was configured with a regularization parameter C=10 and gamma='scale', and trained on the same stratified 80/20 data split used for the LSTM. This setup is representative of traditional non-temporal approaches to SER and provides a meaningful reference point against which the temporal modeling capability of the LSTM can be assessed.

### 2.5 Evaluation Metrics

The performance of the proposed LSTM-based speech emotion recognition model was evaluated using **classification accuracy** and a **confusion matrix**. These metrics were selected to provide both an overall performance measure and a detailed class-level analysis of the model's predictions.

### 2.5.1 Classification Accuracy

The primary evaluation metric that was used to determine the overall correctness of the model was classification accuracy. Accuracy is the number of appropriately labeled samples of the overall number of samples tested. It gives a clear and intuitive as to how the model does with all of the emotion classes. It was ensured that accuracy was kept under training and validation to determine advancement in learning and the probability of overfitting. Since the speech emotion recognition task is multi-class, accuracy would be a good baseline measure of the performance of the model.

### 2.5.2 Confusion Matrix

Besides the general performance of the classification, the use of a confusion table was to further assess the performance of the proposed model per emotion class separately. Although accuracy is a worldwide measure of the rightness, the confusion matrix gives a more in-depth explanation of individualized categorizations and inappropriate classifications. The confusion matrix is used to compare the actual class labels to the predicted labels of the seven types of emotion. The correct prediction values are along the main diagonal and the off-diagonal factor represents the incorrect prediction. Because the problem under consideration is multi-class, the resulting confusion matrix is rather large, but it allows properly assessing the quality of recognition of all the classes of emotions. The confusion matrix was created at the end of the model training with the help of the validation dataset and could be considered an auxiliary measure of evaluation with accuracy.

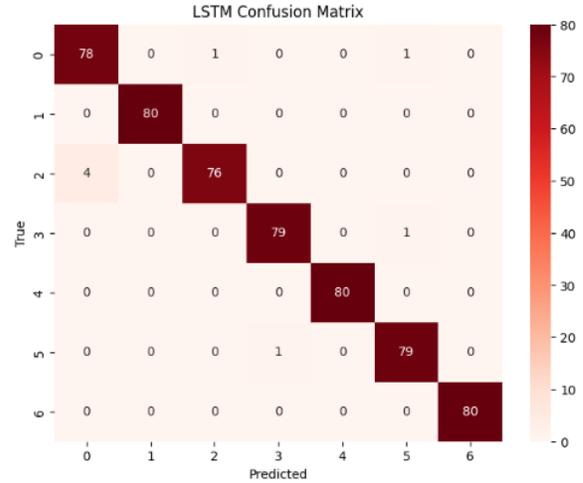

**Fig 20:** Confusion Matrix Visualisation

### 2.5.3 Additional Metrics

While additional metrics such as precision, recall, and F1-score can offer further insights, the combination of **classification accuracy** and **confusion matrix analysis** was considered sufficient for evaluating the proposed model's performance. Overall, the selected evaluation metrics effectively capture both the general and class-specific performance of the proposed model.

## 3. RESULTS

This section presents the performance of the proposed LSTM-based emotion classification model after training for 100 epochs. Model evaluation is based on training and validation accuracy and loss, as well as the final accuracy achieved at convergence.

### 3.1 Training and Validation Performance

Figures **21** and **22** illustrate the training and validation loss and accuracy curves, respectively, across 100 epochs.

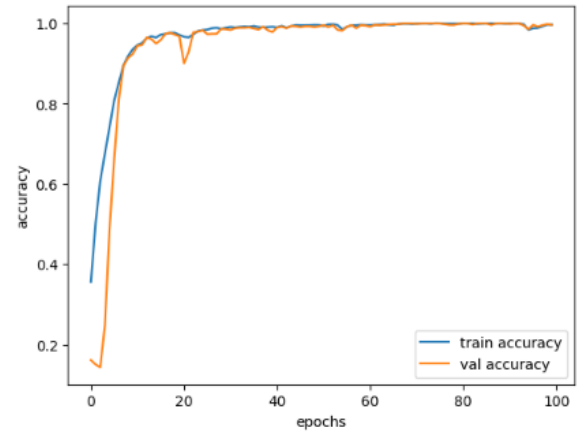

**Fig 21:** Training and validation accuracy versus epochs

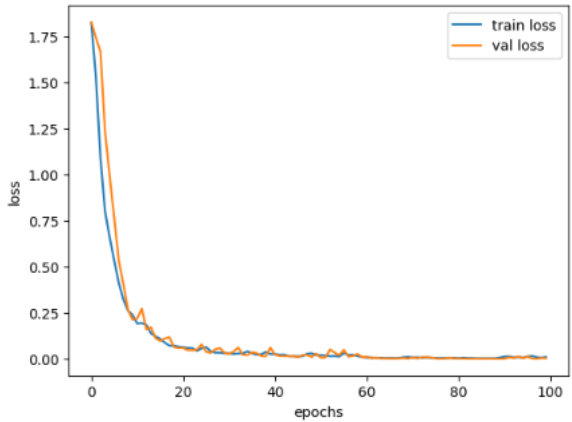

**Fig 22:** Training and validation loss versus epochs

From the loss curve, it can be observed that both the training and validation loss decrease rapidly during the early epochs, indicating fast learning. As training progresses, the loss stabilizes at a very low value, suggesting that the model has successfully learned meaningful representations from the input features. The close alignment between training and validation loss also indicates minimal overfitting.

Similarly, the accuracy curve shows a rapid increase in both training and validation accuracy within the first few epochs. After approximately 20 epochs, accuracy begins to plateau, gradually converging toward near-perfect performance.

### 3.2 Final Epoch Performance

At the end of training (Epoch 100), the model achieved the following results:

- **Training Accuracy: 99.56%**
- **Training Loss: 0.0109**
- **Validation Accuracy: 99.82%**
- **Validation Loss: 0.0036**

These results indicate that the proposed model achieves strong performance on both the training and validation datasets under the current experimental conditions. The low validation loss combined with high validation accuracy suggests that the model is fitting the training data effectively and generalizing well within the scope of this experiment.

### 3.3 Confusion Matrix Analysis

Figure 23 presents the confusion matrix obtained from the trained LSTM model on the validation set. The matrix exhibits strong diagonal dominance, indicating that the substantial majority of samples were correctly classified across all emotion categories.

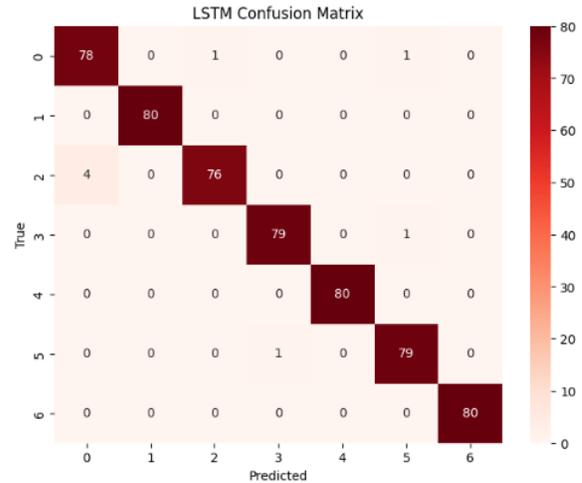

**Fig 23:** Confusion Matrix Analysis

Each emotion class achieved near-perfect classification performance, with most classes recording 80 correct predictions out of 80 samples. Minor misclassifications were observed in only a few cases.

These misclassifications likely occur between emotionally similar classes, which is a common challenge in emotion recognition tasks. Importantly, no class exhibits widespread confusion with multiple other classes, demonstrating strong class separability learned by the model.

Overall, the confusion matrix indicates that the proposed model achieves strong precision and recall across all emotion categories, reinforcing the reliability of the reported accuracy results and validating the robustness of the model for multi-class emotion classification.

The almost-perfect diagonal structure of the confusion matrix further suggests that the model does not suffer from class imbalance or bias toward specific emotion categories.

### 3.4 Summary of Results

Overall, the LSTM-based architecture demonstrates consistent and strong performance across all seven emotion categories on the TESS dataset. The model converges rapidly and remains stable across training and validation, suggesting the selected architecture and training configuration are well suited to this classification task.

### 3.5 Baseline Comparison

Table I presents the test-set classification accuracy of the SVM baseline alongside the proposed LSTM model. The SVM achieved 98% accuracy using static mean MFCC features, confirming that handcrafted acoustic representations carry substantial discriminative information for emotion classification on the TESS dataset. The proposed LSTM, operating on the full temporal MFCC sequence, achieved 99% accuracy on the same test split, suggesting that explicit temporal modeling provides a further performance benefit even in a setting where a classical approach already performs strongly.

**Table I: Model Performance on Test Set**

| Model | Feature Input | Test Accuracy |
|---|---|---|
| SVM (RBF, C=10) [baseline] | Static mean MFCC (40-dim) | 98.00% |
| LSTM (proposed) | Temporal MFCC sequence (40 x T) | 99.00% |

## 4. DISCUSSIONS

The experimental results demonstrate that a single-layer LSTM model trained on MFCC features can achieve very high classification accuracy for speech emotion recognition. This highlights the effectiveness of temporal modeling in capturing emotional patterns embedded in speech signals, even without the use of convolutional or hybrid architectures [2], [13], [16].

### 4.1 Practical Applications

From a practical standpoint, the proposed model is well-suited for applications that require real-time or near-real-time emotion recognition, such as:
- Intelligent virtual assistants and conversational agents
- Mental health and well-being monitoring systems
- Call center analytics and customer sentiment analysis
- Human–computer interaction systems and social robotics

The relatively lightweight architecture, compared to CNN-based or dual-stream models, reduces computational complexity and memory requirements [1], [2]. This suggests potential suitability for real time applications, deployment on resource-constrained devices, including mobile phones and embedded systems, where low latency and efficiency are critical.

### 4.2 Real-World Deployment Considerations

Despite the strong experimental performance, several factors must be considered before deploying the model in real-world environments.

First, audio quality and environmental noise can significantly affect performance. The dataset used in this study consists of relatively clean, well-segmented speech samples. In real deployment scenarios, speech signals may be affected by background noise, reverberation, microphone variability, and overlapping speakers, which could degrade model accuracy.

Second, speaker variability presents another challenge. Differences in age, gender, accent, speaking rate, and emotional expressiveness may not be fully represented in the training data. This can limit the model's ability to generalize to unseen speakers.

Third, latency and system integration must be addressed. Although the model is computationally efficient, real-time deployment would require additional components such as audio streaming, voice activity detection, and preprocessing pipelines, all of which may introduce delays if not carefully optimized.

### 4.3 Limitations of the Proposed Approach

While the results are promising, several limitations must be acknowledged.

#### 4.3.1 Dataset Limitations

The dataset used in this study was obtained from Kaggle and represents a curated collection of labeled emotional speech samples. Although suitable for model development and evaluation, such datasets may not fully capture the diversity of real-world speech. The emotional expressions are often acted or semi-acted, which may differ from spontaneous emotional speech encountered in real-life situations.

Additionally, the dataset size, while adequate for training, may still be insufficient to represent all demographic and cultural variations, raising concerns about bias and generalizability.

### 4.3.2 Model Limitations

The proposed architecture relies solely on MFCC features and a single LSTM layer. While this design choice improves simplicity and efficiency, it may limit the model's ability to capture more complex spectral or prosodic patterns present in emotional speech.

Furthermore, the model was evaluated primarily using accuracy and confusion matrix analysis. Although these metrics indicate strong performance, they may not fully reflect robustness under challenging conditions such as class imbalance or noisy inputs.

### 4.3.3 Generalizability Concerns

The exceptionally high validation accuracy suggests the possibility that the model is particularly well-adapted to the dataset used. This raises concerns about overfitting to dataset-specific characteristics, even though validation performance remains high. Without cross-dataset evaluation, it is difficult to guarantee that similar performance would be achieved on entirely unseen datasets or real-world speech samples.

### 4.4 Future Work

Several directions can be explored to enhance and extend this research.

Future work may involve evaluating the model on multiple datasets to assess generalization and robustness. Incorporating data from different languages, accents, and recording conditions would improve model reliability.

Additional acoustic features such as pitch, energy, and formant-related features could be explored alongside MFCCs to enrich emotional representation. Data augmentation techniques, including noise injection and pitch shifting, may also help improve generalization [9].

From a modeling perspective, experimenting with deeper or bidirectional LSTM architectures, attention mechanisms, or hybrid approaches could further improve performance while maintaining temporal modeling strengths [7], [10], [12].

Finally, future research could focus on real-time system deployment, integrating the model into end-to-end applications and evaluating performance in realistic operational environments.

### 4.5 Conclusion

This paper has explored applying a Long Short-Term Memory network to speech emotion recognition using Mel-Frequency Cepstral Coefficients as the sole acoustic feature. The results indicate that a purely temporal model can achieve strong classification accuracy on the TESS dataset without requiring additional complex structures such as hybrid or convolutional architectures. The training process showed rapid convergence and stable validation performance, with minimal confusion observed between emotion classes. An SVM baseline using static mean MFCC features achieved 98% accuracy, while the proposed LSTM operating on temporal sequences achieved 99%, providing evidence that sequential modeling captures additional discriminative information beyond what static features convey. These findings suggest that streamlined LSTM architectures warrant further investigation as practical candidates for speech emotion recognition tasks, particularly in settings where computational efficiency is a priority. Future work should evaluate generalization across multiple datasets and explore richer feature sets and architectural extensions to assess the broader applicability of this approach.